\documentclass[12pt,reqno]{amsart}
\usepackage{amsmath,amssymb,amsfonts,amsthm}
\usepackage[mathscr]{eucal}
\usepackage[all]{xy}
\usepackage{hyperref}
\usepackage{setspace}
\allowdisplaybreaks

\author{ S.L.~Lyakhovich}
\address{Physics Faculty, Tomsk State University, Lenin ave. 36, Tomsk 634050, Russia.}
\email{sll@phys.tsu.ru} \allowdisplaybreaks

\title{General method for including Stueckelberg fields}

\begin{document}
\maketitle
\begin{abstract}
A systematic procedure is proposed for inclusion of Stueckelberg
fields. The procedure begins with the involutive closure when the
original Lagrangian equations are complemented by all the lower
order consequences. The involutive closure can be viewed as
Lagrangian analogue of complementing constrained Hamiltonian system
with secondary constraints. The involutively closed form of the
field equations allows for explicitly covariant degree of freedom
number count, which is stable with respect to deformations. If the
original Lagrangian equations are not involutive, the involutive
closure will be a non-Lagrangian system. The Stueckelberg fields are
assigned to all the consequences included into the involutive
closure of the Lagrangian system. The iterative procedure is
proposed for constructing the gauge invariant action functional
involving Stueckelberg fields such that Lagrangian equations are
equivalent to the involutive closure of the original theory. The
generators of the Stueckelberg gauge symmetry begin with the
operators generating the closure of original Lagrangian system.
These operators are not assumed to be a generators of gauge symmetry
of any part of the original action, nor are they supposed to form an
on shell integrable distribution. With the most general closure
generators, the consistent Stueckelberg gauge invariant theory is
iteratively constructed, without obstructions at any stage. The
Batalin-Vilkovisky form of inclusion the Stueckelberg fields is
worked out and existence theorem for the Stueckelberg action is
proven.
\end{abstract}

\section{Introduction}
In 1938, Stueckelberg proposed \cite{Stueckelberg} to reformulate
the Proca action for the massive vector field in a gauge invariant
way by introducing the scalar field. Since then, the general idea
attributed to Stueckelberg has been widely used to equivalently
reformulate the original non-gauge theory in a gauge invariant way
by introducing some extra fields. Historical review of ideas about,
and applications of the Stueckelberg method can be found in the
article \cite{Ruegg:2003ps}.

The most often used scheme of introducing the Stueckelberg fields
follows the pattern of the original work \cite{Stueckelberg}
implying that Lagrangian includes a gauge invariant part, and the
non-gauge invariant part. In the case of massive vector field, the
gauge invariant part is the Maxwell Lagrangian, while the
non-invariant part is the massive term. Then, the finite gauge
transformation defining the symmetry of the invariant part is made
of the fields in the entire Lagrangian. This makes the Lagrangian
depending on the gauge parameters. After that, the gauge parameters
are treated as the Stueckelberg fields. Under this scheme, the
Stueckelberg gauge transformations of the original fields are the
same as in the theory without non-invariant part of the Lagrangian.
In this sense, the broken gauge invariance of the original fields is
restored. Once the Stueckelberg fields are included as the
parameters of gauge transformations of the invariant part of the
action, their own gauge symmetry is defined as the composition of
original gauge transformations. In this way, the gauge
transformation of Stueckelberg fields compensates the change of the
non-invariant part of the Lagrangian caused by the transformation of
original fields. The equivalence to the original theory is
established by imposing the gauge conditions fixing the Stueckelberg
fields to zero. In this gauge, the Stueckelberg theory reduces to
the original one. This pattern of inclusion the Stueckelberg fields,
sometimes referred to as the ``Stueckelberg trick", is summarized
and studied in very general form in the article
\cite{Boulanger:2018dau} where one can also find a review of the
vast contemporary literature on this topic.

The ``Stueckelberg trick" works well in many models, but it does not
seem consistent as a general method as it is more art than science.
The division of the Lagrangian into invariant and non-invariant
parts is rather arbitrary, as is the predefined choice of gauge
transformations. For example, given any transformations of the
fields that form the Lie group,  any invariant of the group can be
added to and subtracted from any Lagrangian. Thus, one can get a
division into invariant and non-invariant parts with respect to any
transformation. Thereafter, the above pattern can be applied,
resulting in a model with almost any gauge symmetry that is not
necessarily relevant to the original dynamics. Even the number of
gauge parameters can be any within this approach. In the case of the
Proca action, one could shift the original vector field not by a
gradient of a scalar, but -- for example -- by a divergence of
anti-symmetric second rank tensor. In this case, another part of the
Lagrangian -- the square of divergence of original vector field --
is considered as an invariant, and the rest as a non-invariant part.
The method works, and results in the gauge invariant theory of
massive spin one with reducible gauge symmetry parameterized by
antisymmetric tensor.

Another commonly used scheme of inclusion of Stueckelberg fields is
the method of conversion of the Hamiltonian second class constrained
systems into the first class ones. The first version of the method
\cite{Faddeev:1986pc}, \cite{Batalin:1986aq}, \cite{Batalin:1986fm}
implied to extend the phase space of the original system by new
canonical variables whose number coincides with the number of second
class constraints. After that, the constraints and Hamiltonian are
continued into the extended space  to convert the system into the
first class. Once the gauge conditions are imposed killing the
conversion variables, the system reduces to the original second
class system. The conversion variables can be viewed as Hamiltonian
version of Stueckelberg fields. The original proposal for conversion
implied to linearly include the conversion variables into the
effective first class constraints. In general, the gauge symmetry,
being generated by the effective first class constraints, is
non-abelian. Later on, the abelian conversion scheme has been
proposed \cite{Egorian:1988ss}, \cite{Egorian:1993sc},
\cite{Batalin:1991jm}. Under this scheme, the original second class
constraints are extended by a power series in the conversion
variables to become abelian first class constraints. Any original
phase space function, including Hamiltonian, is extended by the
conversion variables to Poisson commute with the effective abelian
first class constraints. The existence theorem of the abelian
conversion is proven by the homological perturbation theory (HPT)
tools in the article \cite{Batalin:1991jm}. In the article
\cite{Batalin:2005df}, the conversion method is extended to the
general second class systems on symplectic manifolds, with the
constraints not necessarily being scalar functions, but sections of
a bundle over the symplectic base. In this setup, the conversion is
proven to exist, though not necessarily abelian. This conversion
scheme allows one to extend Fedosov deformation quantisation to the
second class constrained systems.

Notice important distinctions of the Hamiltonian conversion method
from the described above ``Stueckelberg trick" which is widely
applied in Lagrangian formalism. The starting point of the
Hamiltonian conversion is a complete system of the second class
constraints, including primary and secondary ones. The Hamiltonian
equations are first order, while primary constraints are zero order.
All these equations are variational. The secondary constraints are
zero order differential consequences of the variational equations,
and they are not variational by themselves. It is the completion of
the original equations by the lower order consequences which allows
one to explicitly count  the degree of freedom (DoF) number. The
same applies not only to the constrained Hamiltonian equations, but
to any system of field equations. The completion of the system by
the lower order consequences is known as the invloutive closure.
Given the involutive closure of the equations, one can count degree
of freedom number in a covariant manner, not appealing to the $1+
(d-1)$ decomposition. Simple and explicitly  covariant DoF number
counting recipe is worked out in  the article \cite{Involution} for
any involutively closed system of field equations, not necessarily
Lagrangian. For the involutive closure of Lagrangian system, the
general recipe (see relation (8) of the article \cite{Involution})
can be further simplified. The covariant DoF count is explained in
the Appendix of this article.  The Hamiltonian scheme of including
Stueckelberg fields proceeds from involutive closure of variational
equations, in particular the number of conversion variables
coincides with the number of constraints. The Lagrangian pattern of
the Stueckelberg trick does not account for the structure of
involutive closure, even the number of the Stueckelberg fields is
unrelated to the number of lower order consequences of the
Lagrangian equations. Notice one more essential distinction between
the pattern of Lagrangian ``Stueckelberg trick" and the Hamiltonian
conversion procedure. The first one proceeds from certain integrable
distribution on the space of fields, which is considered as gauge
symmetry of ``invariant part" of the action. The second class
constraints are unrelated, in general, to any integrable
distribution, and the Hamiltonian conversion methods do not employ
any predefined transformations of the original fields.

One of the motivations for introducing the Stueckelberg fields is
the idea to provide consistent inclusion of interactions by
controlling compatibility of Stueckelberg gauge symmetry when the
free theory is deformed. This idea works well in various examples,
see \cite{Boulanger:2018dau} and references therein. However, it
does not seem a consistent general scheme, as it controls just
algebraic consistency of the Stueckelberg symmetry, not the number
of propagating DoF's, while the artificial symmetry is not
necessarily reasonably related to the structure of the dynamics. In
the article \cite{Involution}, the method is proposed to
consistently include interactions proceeding from the involutive
closure of field equations and without introducing Stueckelberg
fields. Proceeding from the free field equations brought to the
involutive form, the method allows one to iteratively find all the
consistent vertices in the equations. Even though the original
non-involutive equations are Lagrangian, the involutive closure is
not Lagrangian system. Therefore, not all the vertices are
necessarily Lagrangian. For applications of this scheme of inclusion
of interactions, see, for example \cite{Involution},
\cite{Cortese:2013lda}, \cite{Kulaxizi:2014yxa},
\cite{Abakumova:2018eck}, \cite{Rahman:2020qal}.  While
non-Lagrangian vertices may have their own advantages, in particular
they can be stable in higher derivative field theories
\cite{Kaparulin:2014vpa}, \cite{Abakumova:2018eck}, it seems
interesting to have a method of identifying all the consistent
Lagrangian vertices. The way to construct all the consistent
Lagrangian vertices is briefly noticed in the next section as a side
remark.

The main subject of this article  is to work out a method  of
inclusion the Stueckelberg fields which proceeds from the involutive
closure of Lagrangian equations. In this sense, the method can be
considered as the Lagrangian counterpart of the conversion method
for the Hamiltonian second class constrained systems.

\subsection*{Notation} DeWitt's condensed notation
\cite{DeWitt:1965jb} is adopted, where the indices cover both space
time-points $x$ and a set of numerical labels. As a rule, all the
indices are understood as condensed, the exceptions are clear from
the context. The derivatives $\partial_i$ are variational w.r.t. the
fields fields $\phi^i$. Summation over condensed indices includes
integration over $x$. Sign $\approx$ is the on shell equality.

\section{Involutive closure of the Lagrangian equations  and the Stueckelberg fields}
\noindent  In condensed notation, given  the action $S(\phi)$,
Lagrangian equations read
\begin{equation}\label{EoM}
\partial_i S(\phi)=0 \, .
\end{equation}
Suppose the action admits no gauge symmetry. This means that matrix
of second derivatives of the action  does not have on-shell kernel.
Inclusion of Stueckelberg fields in the case with gauge symmetry can
be considered along the similar lines, though it is slightly more
complex. It will be addressed elsewhere.

Let us complement equations (\ref{EoM}) with the consequences
\begin{equation}\label{tau}
   \tau_\alpha(\phi)=- \Gamma^i_\alpha(\phi)\partial_i S(\phi) \, ,
\end{equation}
such that the system
\begin{equation}\label{Closure}
\partial_i S(\phi)=0 \, , \qquad \tau_\alpha =0 \,
\end{equation}
is involutive. Here, the involution means that the system does not
admit any lower order consequence which is not already included. For
a review of the involution concept in the partial differential
equation (PDE) theory, and various applications, one can consult the
book \cite{Seiler}. If the original Lagrangian equations are not
involutive, by adding consequences, it will be brought to the
involution. This can be considered as Lagrangian analogue of
Dirac-Bergmann algorithm of constrained Hamiltonian formalism with
the consequences $\tau_a$ being analogues of the second class
constraints. The difference is that the order of derivatives is
considered w.r.t. any space-time coordinate. Much like the
Hamiltonian conversion, the Lagrangian procedure of inclusion
Stueckelberg fields begins with involutive closure of equations of
motion.

In principle, one can include the consequences of the higher order
than the original Lagrangian equations. The only requirement is that
the system of Lagrangian equations and their consequences
(\ref{Closure}) are involutive --- i.e. any lower order consequence
is already contained among these equations\footnote{ This also has a
counterpart in the constrained Hamiltonian formalism. Given the
Lagrangian with the first order derivatives, one can construct the
Hamiltonian formalism as if the acceleration were included. Then,
the phase space would include auxiliary coordinates, absorbing
velocities, and also extra momenta. These extra variables are
suppressed by the second class constraints. In principle, these
constraints can be converted into the first class, by usual
conversion procedure. This conversion procedure, at Lagrangian
level, would correspond to the involutively closed system
constructed by inclusion of the higher order consequences of
original Lagrangian equations. }. Example of this sort is considered
in the end of this section.

 The consequences $\tau_\alpha$ are supposed independent. This also
 means that the generators $\Gamma^i_\alpha$, being the local
 differential operators, are also independent.
 The specific conditions of independence are explained in the next
 section. The over-complete set of generators $\Gamma$ would lead to
 the reducible Stueckelberg gauge symmetry. This case is not considered
 in the article, though it can be of interest for some models.

The involutive system (\ref{Closure}) admits gauge identities
\begin{equation}\label{GI}
      \Gamma^i_\alpha(\phi)\partial_i S(\phi) + \tau_\alpha(\phi)\equiv 0 \,
\end{equation}
which can be viewed as a rephrasing of the definition of
consequences $\tau_\alpha$ (\ref{tau}). The involutively closed
system (\ref{Closure})  is non-Lagrangian as such though it is
equivalent to the original Lagrangian system (\ref{EoM}). In
non-Lagrangian systems, the second Noether theorem does not apply,
so the gauge identities are not necessarily related to a gauge
symmetry. Since the original Lagrangian equations (\ref{EoM}) do not
have gauge symmetry, then their involutive closure (\ref{Closure})
will not be gauge invariant either, because they define the same
mass shell.

The involutive closure (\ref{Closure}) of the Lagrangian system is
characterised by three types of numbers which determine the DoF
number: (i) the orders of original Lagrangian equations (\ref{EoM});
(ii) the orders of the consequences $\tau_a$ included into
involutive closure (\ref{Closure})of the system; (iii) the orders of
differential operators $\Gamma^i_\alpha$ generating consequences of
Lagrangian equations included into the involutive closure. The DoF
number is a certain linear combination of these three types of
integers. The DoF counting is detailed in the Appendix, see relation
(\ref{DoF1}) and corresponding explanations.

 Let us make a side remark on consistent inclusion of
interactions, proceeding from the involutive closure (\ref{Closure})
of the Lagrangian system. Given the free Lagrangian without gauge
symmetry, the problem is to find all the vertices such that the DoF
number remains unchanged upon inclusion of interactions. The
procedure is quite simple. At first, one has to bring the system of
the free Lagrangian equations into the involutive form, by
complementing them with all the lower order consequences. Given the
involtutive closure, the DoF number is fixed. The second step to
consistent inclusion of interactions is to \emph{simultaneously}
deform the free action $S$ \emph{and} the generators
$\Gamma_\alpha^i$ of the consequences,
\begin{equation}\label{Vertices}
    S(\phi)=\frac12M_{ij}\phi^i\phi^j\,\mapsto\, \textsf{S}_{int}=\sum_{k=0}\stackrel{(k)}{\textsf{S}}(\phi)\,, \qquad
     \stackrel{(k)}{\textsf{S}}= \frac{1}{k+2} M_{i_1\dots
     i_{k+2}}\phi^{i_1}\dots \phi^{i_{1+2}}\, ;
\end{equation}
\begin{equation}\label{Gamma-vertices}
    \Gamma^i_\alpha \, \mapsto \, \textsf{G}^i_\alpha
    (\phi)=\sum_{k=0}\stackrel{(k)}{\textsf{G}}{}^i_\alpha\, ,
    \qquad
    \stackrel{(k)}{\textsf{G}}{}^i_\alpha=G^i_\alpha{}_{j_1\dots
    j_k}\phi^{j_1}\dots\phi^{j_k} \, ,
\end{equation}
where the vertices $M_{i_1\dots i_{k+2}},\, G^i_\alpha{}_{j_1\dots
j_k} $ are field-independent poly-differential operators. The first
operator $M_{ij}$  is the same as in the free theory, and
$\stackrel{(0)}{\textsf{G}}{}^i_\alpha $ coincides with the
generator of the consequences included into the involutive closure
at the free level.
 Given  the vertices in the action (\ref{Vertices}) and the generators of consequences (\ref{Gamma-vertices}),
 one gets the deformation of the consequences included into the involutive
 closure (\ref{Closure})
 \begin{equation}\label{tau-vertices}
    \tau_\alpha= -\Gamma^i_\alpha\partial_iS\,\,\mapsto\,\, - \textsf{G}^i_\alpha\partial_i
    \textsf{S}_{int}=  \textsf{T}_\alpha=\sum_{k=0}\stackrel{(k)}{\textsf{T}}_\alpha\,
    \, , \qquad \stackrel{(k)}{\textsf{T}}_\alpha= \textsf{T}_{\alpha i_1\dots i_{k+1}}\phi^{i_1}\dots \phi^{i_{k+1}} \, ;
\end{equation}
\begin{equation}\label{Tk}
\textsf{T}_{\alpha i_1\dots
i_{k+1}}=\sum_{l=0}^{k+1}G^i_\alpha{}_{(j_1\dots
    j_l}M_{j_{l+1}\dots j_{k+1})i} .
\end{equation}
Here, the round brackets mean symmetrization of corresponding
indices. Upon inclusion of interaction, the deformed action
(\ref{Vertices}) and consequences (\ref{tau-vertices}) by
construction obey the gauge identity:
\begin{equation}\label{ID-int}
\textsf{G}^i_\alpha\partial_i
    \textsf{S}_{int}+ \textsf{T}_\alpha\equiv 0 .
\end{equation}
One can see once again that from the perspective of algebraic
consistency, any interaction is admissible for the field equations
without gauge symmetry. The consistency of interactions is provided
not just by algebraic consistency but also the DoF number should
remain unchanged upon deformation. This condition can be easily
controlled making use of the involtive form of field equations. The
vertices (\ref{Vertices}), (\ref{Gamma-vertices}) will be consistent
if the following two conditions are met: (i) the system
\begin{equation}\label{Int-Closure}
\partial_i\textsf{S}_{int} =0\,,\qquad \textsf{T}_\alpha = 0
\end{equation}
remains involutive; (ii) the DoF number (\ref{DoF1}) for equations
(\ref{Int-Closure}) remains the same as it is for the free system.
Upon inclusion of interaction, the orders of Lagrangian equations,
generators, and consequences, being ingredients of the DoF number
count, can increase, or remain unchanged. If they do not increase,
the DoF number obviously remains unchanged. These three orders can
increase, however, in the correlated way without changing the DoF
number, once relation (\ref{DoF1}) still holds true. Therefore, the
interaction can be consistent, in principle, even if the higher
derivatives are involved.

Below in this section, the iterative procedure is described for
inclusion of Stueckelberg fields. Under this procedure, the
Stueckelberg field $\xi^\alpha$ is assigned to every consequence
$\tau_\alpha (\phi)$ included into the involutive closure
(\ref{Closure}) of original Lagrangian equations. The Stueckelberg
action is sought for as a power series in the fields $\xi^\alpha$:
\begin{equation}\label{St-action}
  \mathcal{S}_{St}(\phi,\xi)=   \sum_{k=0}^\infty \mathcal{S}_k\, ,\qquad \mathcal{S}_k(\phi,\xi)=  W_{\alpha_1\ldots\alpha_k}(\phi)\, \xi^{\alpha_1} \cdots
  \xi^{\alpha_k}\,, \quad k>0 \,   ,
\end{equation}
where $\mathcal{S}_0(\phi)$ is the original action $S(\phi)$, and
the first expansion coefficient $W_\alpha$ is defined by
$\tau_\alpha$ (\ref{tau}):
\begin{equation}\label{tau+}
    W_\alpha(\phi)=\frac{\partial \mathcal{S}_{St}(\phi,\xi)}{\partial
    \xi^\alpha}{}_{\displaystyle{\big|\xi=0}}=
    \tau_\alpha \, .
\end{equation}
Hence, at $\xi=0$, the field equations for the Stueckelberg action
reproduce the involutive closure (\ref{Closure}) of the original
Lagrangian equations.

The equivalence of the Stueckelberg theory  to the original one is
provided by the gauge symmetry of the action (\ref{St-action}) such
that the fields $\xi^\alpha$ can be gauged out, with $\xi^\alpha=0$
being admissible gauge fixing condition. This means, the number of
gauge parameters should coincide  with the number of consequences
$\tau_\alpha$ (\ref{tau}) included into involutive closure of the
original Lagrangian system. The gauge transformations are
iteratively sought for order by order of the Stueckelberg fields
\begin{equation}\label{gt}
    \delta_\epsilon\phi^i=R^i_\alpha(\phi,\xi)\epsilon^\alpha\, ,
    \quad \delta_\epsilon\xi^\gamma= R^\gamma_\alpha(\phi,\xi)\epsilon^\alpha\,
    , \quad
    R_\alpha^i (\phi,\xi)=
    \sum_{k=0} \stackrel{(k)}{R}{}^i_\alpha, \quad R_\alpha^\gamma (\phi,\xi)=
    \sum_{k=0} \stackrel{(k)}{R}{}^\gamma_\alpha,
\end{equation}\label{Rk}
\begin{equation}
   \stackrel{(k)}{R}{}_\alpha^i (\phi,\xi)=
    R^i_{\alpha\beta_1\dots\beta_k}(\phi)\xi^{\beta_1}\dots\xi^{\beta_k}\,, \qquad  \stackrel{(k)}{R}{}_\alpha^\gamma (\phi,\xi)=
    R^\gamma_{\alpha\beta_1\dots\beta_k}(\phi)\xi^{\beta_1}\dots\xi^{\beta_k}\,.
\end{equation}
The Stueckelberg action (\ref{St-action}) is supposed to be
invariant with respect to the gauge transformations (\ref{gt}).
 The gauge symmetry (\ref{gt}) of the action is equivalent to the Noether
identities
\begin{equation}\label{GSSt}
    \delta_\epsilon\mathcal{S}_{St}= 0\,, \quad
    \forall\epsilon^\alpha\quad\Leftrightarrow\quad
    R^i_\alpha\partial_i\mathcal{S}_{St} +R^\gamma_\alpha \frac{\partial\mathcal{S}_{St} }{\partial
   \xi^\gamma} \equiv 0\, .
\end{equation}
These identities  can be expanded in Stueckelberg fields,
\begin{equation}\label{NI-exp}
   \left( R_\alpha^\gamma(\phi,\xi)\frac{\partial}{\partial
   \xi^\gamma}+ R_\alpha^i(\phi,\xi)\frac{\partial}{\partial
   \phi^i}\right)\mathcal{S}_{St}\equiv \sum_{k=0}
   \sum_{m=0}^k\left(\stackrel{(k-m)}{R}{}_\alpha^\gamma \,
\frac{\partial \mathcal{S}_{(m+1)}}{\partial\xi^\gamma}\, + \,
\stackrel{(k-m)}{R}{}_\alpha^i \, \frac{\partial
\mathcal{S}_{(m)}}{\partial\phi^i}\right) \equiv 0 \, .
\end{equation}
Once the Noether identities are valid for every order in $\xi$, each
term in the sum over $k$ should vanish separately. In this way, the
requirement of the gauge symmetry results in the sequence of
relations
\begin{equation}\label{NI-k}
\sum_{m=0}^k\left(\stackrel{(k-m)}{R}{}_\alpha^\gamma \,
\frac{\partial \mathcal{S}_{(m+1)}}{\partial\xi^\gamma} \,\, + \,\,
\stackrel{(k-m)}{R}{}_\alpha^i \, \frac{\partial
\mathcal{S}_{(m)}}{\partial\phi^i}\right) \equiv 0 \,  , \quad
k=0,1, 2,\dots\, .
\end{equation}
For $k=0$, given the boundary condition (\ref{tau+}),  the above
relations reduce to the identities between consequences $\tau$ and
Lagrangian equations,
\begin{equation}\label{W1}
\stackrel{(0)}{R}{}^\gamma_\alpha\tau_\gamma +
\stackrel{(0)}{R}{}^i_\alpha\partial_iS\equiv 0 \, .
\end{equation}
Any identity between $\tau_\alpha$ and $\partial_iS$
reduces\footnote{Accurate formulation  of completeness of the
identities (\ref{GI}) is provided in the next section, see in
particular relations (\ref{completeGI}), (\ref{U}). Here we proceed
from the intuitive understanding that the consequences $\tau$, being
defined as \emph{independent} linear combinations of the equations
by relations (\ref{tau}), cannot admit any other dependency with
$\partial_iS$ besides the one following from the definition. } to
the linear combination of the identities in the closure of the
original system (\ref{GI}). This means, the identity (\ref{W1})
reads
\begin{equation}\label{R0-Id}
\stackrel{(0)}{R}{}^\gamma_\alpha\left(\tau_\gamma +
\Gamma^i_\gamma\partial_iS\right)\equiv 0 \, ,
\end{equation}
where $\stackrel{(0)}{R}{}^\gamma_\alpha$ can be any non-degenerate
matrix. Below, we stick to the simplest choice
\begin{equation}\label{R0-delta}
\stackrel{(0)}{R}{}^\gamma_\alpha= \delta^\gamma_\alpha\, .
\end{equation}
This choice does not restrict the generality for two reasons. First,
as demonstrated below, it admits consistent inclusion of
Stueckelberg fields for any action and any generating set of the
consequences included into the involutive closure of original
Lagrangian system (\ref{Closure}). Second, any other choice of
non-degenerate $\stackrel{(0)}{R}{}^\gamma_\alpha$ can be absorbed
by the change of the gauge parameters $\epsilon^\alpha$.

Given relation (\ref{R0-Id}), the choice (\ref{R0-delta})  defines
zero order of the Stueckelber gauge symmetry (\ref{gt}) for the
original fields:
\begin{equation}\label{R0-Gamma}
\stackrel{(0)}{R}{}^i_\alpha= \Gamma^i_\alpha \, .
\end{equation}
Given zero order of the expansion for the gauge transformations
(\ref{gt}) in $\xi$, and zero and first order in the Stuekelberg
action (\ref{St-action}),
\begin{equation}\label{first-approx}
\mathcal{S}_{St}(\phi,\xi)=S(\phi)+\tau_\alpha(\phi)\xi^\alpha+
\dots\, , \quad \delta_\epsilon\phi^i= \Gamma^i_\alpha (\phi)
\epsilon^\alpha+\dots\, ,
\quad\delta_\epsilon\xi^\alpha=\epsilon^\alpha+\dots\, ,
\end{equation}
all the higher orders are iteratively defined by relations
(\ref{NI-k}), both for the action, and for the gauge
transformations. The procedure of resolving relation (\ref{NI-k})
with certain $k$ for $\mathcal{S}_{(k+1)}$ and
$\stackrel{(k)}{R}_\alpha$ is inductive. Relations (\ref{W1}),
(\ref{R0-delta}), (\ref{R0-Gamma}) solve equations (\ref{NI-k}) for
$k=0$. Substituting this solution into ((\ref{NI-k}) with $k=1$, we
get the equation for $\mathcal{S}_{(2)}$ and
$\stackrel{(1)}{R}_\alpha$. This equation is labeled by index
$\alpha$, and it is linear in $\xi^\beta$. Once the relation has to
be met for any $\xi^\beta$,  the equation is a square matrix. The
symmetric part of the matrix defines the structure coefficient
$W_{\alpha\beta}$ of $\mathcal{S}_{(2)}$, while the anti-symmetric
part defines the structure coefficients of
$\stackrel{(1)}{R}_\alpha$. The solution involves certain ambiguity
related to the fact that gauge generators are defined modulo
on-shell vanishing contributions, and up to a linear combination. It
is the ambiguity which is common for any gauge theory. Given
$\stackrel{(l)}{R}_\alpha, \, \mathcal{S}_{(l+1)}, \,
l=0,1,\dots,k$, they all are substituted into equation (\ref{NI-k})
of the order $k+1$ that defines $\stackrel{(k+1)}{R}_\alpha$ and
$\mathcal{S}_{(k+2)}$. This iterative procedure is unobstructed at
any stage. This is seen from the algebraic consequences of the gauge
identity studied in the next section. The formal proof of
consistency of the procedure for inclusion the Stueckelberg feilds
is provided in Section 4 by the HPT tools.

\section{Gauge algebra of the involutive closure of Lagrangian system}
In this article, the class of field theories is considered such that
the action does not admit gauge symmetry\footnote{Inclusion of
Stueckelberg fields in the models enjoying another gauge symmetry
can be considered along the same lines, though the procedure would
need some adjustments.}, while the Lagrangian equations are not
involutive. The involutive closure (\ref{Closure})
 is non-Lagrangian as such, though it is equivalent to the original
 Lagrangian system (\ref{EoM}). Since the original equations
 (\ref{EoM}) are complemented by their consequences $\tau_\alpha$ (\ref{tau}),
 the extended system (\ref{Closure}) admits gauge identities
 (\ref{GI}). These identities are unrelated to any gauge symmetry.
 This is possible because the involutively closed system (\ref{Closure})
 is non-Lagrangian, so the second Noether theorem does not apply.
 The procedure of inclusion Stueckelberg fields
 described in the previous section is intended to convert the
 identities (\ref{GI}) into Noetherian ones by introducing the new fields $\xi$
 and gauge symmetry such that the generators of the identities are
 converted into  gauge generators (\ref{first-approx}).

The gauge identities (\ref{GI}) turn out having a sequence of
consequences of their own. The gauge identities and their
consequences are  understood as the gauge algebra of the involutive
closure (\ref{Closure}). It is the algebra which is behind existence
of the solution to the conversion equations (\ref{NI-k}) in every
order. This algebra is considered in this section.

Let us begin with the remarks concerning equivalence relations for
the generators of consequences $\Gamma^i_\alpha$ (\ref{tau}). First,
notice that the original action $S(\phi)$ is assumed having no gauge
symmetry. Hence, if an identity occurs between Lagrangian equations,
the identity generator is trivial,
\begin{equation}\label{TrivL}
    \kappa^i\partial_iS\equiv 0\quad\Leftrightarrow\quad \exists E^{ij}=-E^{ji}:
    \,\,\kappa^i=E^{ij}\partial_j S \, .
\end{equation}
Since the consequences $\tau_\alpha$ (\ref{tau}) are supposed
independent, any identity between them is trivial in the similar
sense
\begin{equation}\label{Ind-tau}
    \kappa^\alpha \tau_\alpha \equiv 0\quad\Leftrightarrow\quad\exists E^{\alpha\beta}=-E^{\beta\alpha}: \,  \kappa^\alpha = E^{\alpha\beta}\tau_\beta \, .
\end{equation}
The generators of the of invloutive closure  $\Gamma^i_\alpha$
(\ref{tau}) are considered equivalent if they result in the same
consequences $\tau_\alpha$. Hence, the difference between equivalent
generators $\Gamma_\alpha^i$  and $\Gamma'{}_\alpha^i$ is a trivial
generator of the identity between the original equations:
\begin{equation}\label{GG'}
    \Gamma^i_\alpha(\phi)\partial_i S(\phi)-\Gamma'{}^i_\alpha(\phi)\partial_i S(\phi)\equiv 0 \quad\Leftrightarrow\quad \Gamma^i_\alpha-\Gamma'{}^i_\alpha = E_\alpha^{ij}\partial_j S, \quad  E_\alpha^{ij}=- E_\alpha^{ji} .
\end{equation}
Once the consequences $\tau_\alpha$ (\ref{tau}) are independent, any
set of identities (we label the set elements by the index $A$) among
the involutive equations (\ref{Closure}) is spanned by the
identities (\ref{GI}):
\begin{equation}\label{completeGI}
    \Lambda_A^i\partial_iS+\Lambda^\alpha_A\tau_\alpha\equiv 0
    \quad\Leftrightarrow\quad \exists U^\alpha_A: \,
    \Lambda_A^i\partial_iS+\Lambda^\alpha_A\tau_\alpha\equiv
    U^\alpha_A\left(\Gamma^i\partial_i S +\tau_\alpha\right) \, .
\end{equation}
The expansion coefficients $U_A^a$ define the generators of the
identities $\Lambda_A$ modulo natural ambiguity
\begin{equation}\label{U}
    \Lambda_A^i=U^\alpha_A\Gamma_\alpha^i +E^{ij}_A\partial_j S +
       E^{i\alpha}_A \tau_\alpha \, ,\quad \Lambda_A^\alpha=
     U^\alpha_A - E^{i\alpha}_A\partial_i S +
     E^{\alpha\beta}_A\tau_\beta\,, \quad E^{ij}_A=-E^{ji}_A\,
     ,\quad E^{\alpha\beta}_A=-E^{\beta\alpha}_A,
\end{equation}
with $E_A$ being arbitrary. All the relations above are valid, in
principle, for any regular system of field equations admitting
irreducible generating set for gauge identities. These relations do
not imply that the equations (\ref{Closure}) follow from Lagrangian
equations (\ref{EoM}).

Now, let us exploit the fact that the original field equations are
Lagrangian to deduce the consequences of the gauge identities
(\ref{GI}). Consider the action of variational vector field
$\Gamma_\alpha=\Gamma^j_\alpha\partial_j$ onto the consequence
$\tau_\beta$ (\ref{tau}). On shell, this amounts to the second
variation of the action $S$ along the field $\Gamma_\alpha$. Since
the variational derivatives commute, the second variations of the
action along variational vector fields commute on shell
\begin{equation}\label{Gtau}
    \Gamma^i_\alpha(\phi)\partial_i \tau_\beta(\phi)\approx\Gamma_\alpha^i\Gamma_\beta^j \partial^2_{ij} S = \Gamma^i_\beta\Gamma^j_\alpha\partial^2_{ij}S \, .
\end{equation}
 Once the matrix $\Gamma^i_\alpha(\phi)\partial_i \tau_\beta$
 is symmetric on shell, off shell the symmetry can be broken by the contributions proportional to $\tau_\alpha$ and $\partial_i S$:
\begin{equation}\label{W}
\Gamma^i_\alpha(\phi)\partial_i \tau_\beta = W_{\alpha\beta} + R^i_{\alpha
    \beta}\partial_i S + R^\gamma_{\alpha
    \beta}\tau_\gamma\,  , \quad W_{\alpha\beta}= W_{\beta\alpha}.
\end{equation}
The matrix $W_{\alpha\beta}$ is off shell symmetric. On shell,
$W_{\alpha\beta}\approx\Gamma_\alpha^i\Gamma_\beta^j \partial^2_{ij}
S $. The structure coefficients $R^\gamma_{\alpha\beta},
R^i_{\alpha\beta}$ do not have certain symmetry w.r.t. the lower
labels. Consider antisymmetric part of relations (\ref{W}), and use
the definition of $\tau_\alpha$ (\ref{tau})
\begin{equation}\label{GR}
\left(\Gamma^i_{[\alpha}(\phi)\partial_i \Gamma_{\beta]}^j
-R^j_{[\alpha
    \beta]}\right)\partial_jS - R^\gamma_{[\alpha
    \beta]}\tau_\gamma\equiv 0\,.
\end{equation}
The relation above is the identity between the original Lagrangian
equations $\partial_i S$ and their consequences $\tau_\alpha$. Any
identity of this type reduces to the basic  identity (\ref{GI})
according to relation  (\ref{completeGI}). The coefficients  in this
identity are connected with each other by relation (\ref{U}).
Applying (\ref{U}) to the specific identity (\ref{GR}) we arrive at
the following relations defining the set of structure functions
$R^i_{\alpha\beta}, R^\gamma_{\alpha\beta}$ involved in (\ref{GR})
in terms of a single independent structure coefficient
$U^\gamma_{\alpha\beta}$ and arbitrary structure functions $E$:
\begin{equation}\label{GG}\Gamma^j_\alpha\partial_j \Gamma_\beta^i-\Gamma^j_\beta\partial_j
\Gamma_\alpha^i-U^\gamma_{\alpha\beta}\Gamma^i_\gamma-R^i_{\alpha
    \beta}+R^i_{\beta\alpha}-E^{ji}_{\alpha\beta}\partial_j
    S-E^{i\gamma}_{\alpha\beta}\tau_\gamma = 0;
\end{equation}
\begin{equation}\label{UR}
U^\mu_{\alpha\beta} - R^\mu_{\alpha
    \beta}+R^\mu_{\beta \alpha}+E^{j\mu}_{\alpha\beta}\partial_j
    S-E^{\mu\nu}_{\alpha\beta}\tau_\nu=0 \, .
\end{equation}
Let us briefly comment on relations (\ref{GG}), (\ref{UR}). The
first of them demonstrates that the generators of the consequences
$\Gamma^i_\alpha$ do not necessarily form an on-shell integrable
distribution, unlike the generators of gauge symmetry. These
generators commute on shell to their linear combinations with the
structure coefficients $U_{\alpha\beta}^\gamma$, while
$R^i_{\alpha\beta}$ describes deviation from integrability of the
distribution. Both $U_{\alpha\beta}^\gamma$ and $R^i_{\alpha\beta}$
are defined modulo natural off-shell ambiguity, given the original
generators $\Gamma_\alpha^i$. Relations (\ref{UR}) identify the
anti-symmetric part of structure function $R_{\alpha\beta}^\gamma$
involved in the relation (\ref{Gtau}) as the involution coefficient
$U_{\alpha\beta}^\gamma$ in commutation relations of generators
(\ref{GG}). Symmetric parts of $R_{\alpha\beta}^\gamma,
R^i_{\alpha\beta}$ can be absorbed by on-shell vanishing part of
symmetric structure function $W_{\alpha\beta}$ defined by relation
(\ref{W}).

Notice that relations (\ref{W}), (\ref{GG}), (\ref{UR}) are the
immediate consequences of the identity (\ref{GI}). They follow  from
the fact that the set of identities (\ref{GI}) is complete and
irreducible. In their turn, the identities (\ref{GI}) follow from
the definition of the functions $\tau$ as independent linear
combinations (\ref{tau}) of the l.h.s. of Lagrangian equations
(\ref{EoM}). Once the original equations are consistent, all their
consequences, including relations (\ref{W}), (\ref{GG}), (\ref{UR})
cannot be inconsistent.

Once the structure coefficients $R_{\alpha\beta}^\gamma,
R^i_{\alpha\beta}, W_{\alpha\beta}$  are found from relations
(\ref{W}), (\ref{GG}), (\ref{UR}) modulo natural ambiguity, they
define the first order of the Stueckelberg gauge symmetry generators
 and the second order of Stueckelberg action,
\begin{equation}\label{R1S2}
\stackrel{(1)}{R}{}^i_\alpha=R^i_{\alpha\beta}\xi^\beta\, , \qquad
\stackrel{(1)}{R}{}^\gamma_\alpha=R_{\alpha\beta}^\gamma\xi^\beta\,,\qquad
\mathcal{S}_{(2)}= \frac12W_{\alpha\beta}\xi^\alpha\xi^\beta \, .
\end{equation}
Notice that if the generators $\Gamma^i_\alpha$ of consequences
included into involutive closure of Lagrangian equations form
integrable distribution, $\stackrel{(1)}{R}{}^i_\alpha$ will vanish,
as $R^i_{\alpha\beta}=0$. In this case, the Stueckelberg symmetry
does not mix up the original fields with the Stueckelberg ones, much
like it happens in the ``Stueckelberg trick". If the distribution
generated by $\Gamma$ is not integrable, i.e. $R^i_{\alpha\beta}\neq
0$, the deviation from integrability is included into the generator
of Stueckelberg symmetry at the first order in $\xi$, compensating
non-commutativity of zero order term.   As demonstrated in the next
section, the iterative procedure consistently continues in the
higher orders, and it inevitably results in the gauge
transformations with on-shell integrable distribution.

Concluding this section, let us mention that relations (\ref{W}),
(\ref{GG}), (\ref{UR}) can have further consequences involving
higher structure functions. These higher structures contribute to
the higher orders in the Stueckelberg action and gauge generators.
All these higher relations should be also consistent as the original
Lagrangian equations are supposed having no contradictions, and
hence any inconsistency is impossible in their consequences.

\section{BV master equation for the Stueckelberg gauge symmetry}
In this section, the BV formalism is rearranged to serve as a tool
for consistent inclusion of Stueckelberg fields.

If the Stueckelberg action $ \mathcal{S}_ {St} (\phi, \xi) $
(\ref{St-action}) and the corresponding gauge generators $R_\alpha^i
(\phi, \xi), R_\alpha^\beta (\phi, \xi) $ (\ref{gt}), (\ref{GSSt})
would have been known from the outset, it could be considered as the
usual Lagrangian gauge theory without any distinction between
Stueckelberg fields $ \xi^\alpha $ and original fields $\phi^i $.
Then, the master action can be constructed for the gauge system
along the usual lines of the BV formalism \cite{BV1}, \cite{BV4}.
The ghosts $C^\alpha$ are assigned to all the gauge parameters
$\epsilon^\alpha$, and the anti-fields are introduced for all
fields, including ghosts. The usual Grassmann parity and ghost
number gradings are imposed on the fields and
antifields\footnote{For simplicity, we consider the case with even
gauge symmetries and even original fields. Adjustments are
 made to the odd case by inserting known sign factors.}:
\begin{equation}\label{grad-fields}
\begin{array}{cccc}
  \epsilon(\phi^i)=\epsilon(\xi^\alpha)=0, &  \epsilon
(C^\alpha)=1, & gh(\phi^i)=gh(\xi^\alpha)=\,0\, ,\,  & gh(C^\alpha)=\phantom{1}1 ; \\
\epsilon(\phi^*_i)=\epsilon(\xi^*_\alpha)=1, & \epsilon
(C^*_\alpha)=0, & gh(\phi^*_i)=gh(\xi^*_\alpha)=-1, & gh(C^*_\alpha)=-2. \\
\end{array}
\end{equation}
The BV action is sought for as a solution to the master equation
\begin{equation}\label{ME}
    (S_{BV}, S_{BV})=0,\qquad gh(S_{BV})=0\,, \quad\epsilon (S)=0,
\end{equation}
where $(\cdot,\cdot)$ is the anti-bracket
\begin{equation}\label{anti-br}
    (A,B)=\frac{\partial^RA}{\partial
    \varphi^I}\frac{\partial^LB}{\partial\varphi^\ast_I}-\frac{\partial^RA}{\partial
    \varphi^\ast_I}\frac{\partial^LB}{\partial\varphi^I}\,,\qquad
    \varphi^I=(\phi^i,\xi^\alpha,C^\alpha)\,,\qquad
    \varphi^\ast_I=(\phi^\ast_i,\xi^\ast_\alpha,C^\ast_\alpha)\,.
\end{equation}
To solve the master equation, the usual setup of the BV formalism
for irreducible gauge systems would imply imposing the boundary
condition on the action,
\begin{equation}\label{BC-usual}
S_{BV}(\varphi,\varphi^*)=\mathcal{S}_{St}(\phi,\xi)+
C^\alpha\left(R^i_\alpha(\phi,\xi)\phi^*_i+
R^\gamma_\alpha(\phi,\xi)\xi^*_\gamma\right)+\dots \,.
\end{equation}
Here, the first term is the gauge invariant action, and the second
one includes generators of the gauge symmetry of the action
multiplied by corresponding ghosts $C^\alpha$ and anti-fields
$\phi^*_i$, $\xi^*_\gamma$. This term has anti-ghost degree 1. The
dots stand for the terms of higher anti-ghost degrees. The
anti-ghost degree is imposed in usual way \cite{Henneaux:1992ig}:
\begin{equation}\label{antigh}
agh(\phi^*_i)=agh(\xi^\alpha)=1, \quad agh(C^*_\alpha)= 2; \quad
agh(\phi^i)=agh(\xi^\alpha)=agh(C^\alpha)=0 \,.
\end{equation}
Given the boundary condition (\ref{BC-usual}), the higher terms can
be iteratively found from the master equation (\ref{ME}) by
expansion  with respect to the anti-ghost degree. The unique
existence of the solution  can be proven by the usual HPT tools
\cite{BV4}, \cite{Henneaux:1992ig}.

From the perspective of including Stueckelberg fields in the BV
formalism, the boundary condition (\ref{BC-usual}) is unsuitable,
because neither the Stueckelberg  action $\mathcal{S}_{St}$ is known
from the outset, nor are the gauge generators $R_\alpha^\beta,
R_\alpha^i$. The action is known up to the first order in
Stueckelberg fields (\ref{tau+}), while the gauge generators are
known only  at $\xi^\alpha=0$, see (\ref{R0-delta}),
(\ref{R0-Gamma}). Hence, the known part of the boundary condition
(\ref{BC-usual}) reads
\begin{equation}\label{BC-St}
  S_{BV}(\varphi,\varphi*) = S(\phi) - \tau_\alpha (\phi)\xi^\alpha + C^\alpha\Gamma^i_\alpha (\phi) \phi_i^* +
  C^\alpha\xi_\alpha^* \,  + \, \dots \, ,
\end{equation}
where the dots stand for the higher orders of anti-fields \emph{and}
Stueckelberg fields. So, to find the solution for the master
equation (\ref{ME}) of the Stueckelberg theory one has to proceed
from the boundary condition (\ref{BC-St}) iterating the solution
order by order w.r.t. anti-fields and Stueckelberg fields. This
means, another resolution degree has to be imposed instead of the
anti-ghost number (\ref{antigh}) such that would be nonzero for
Stueckelberg fields. The boundary condition (\ref{BC-St}) should be,
at maximum, of the first order in the resolution degree, so $\xi$
has to be assigned the weight 1. So, we impose the following
resolution degree:
\begin{equation}\label{deg}
    deg(\xi^\alpha)=deg(\xi^*_\alpha)=deg(\phi^*_i)=1, \quad deg(C^*_\alpha)=2, \quad
    deg(C^\alpha)=deg (\Phi^i)=0 \,.
\end{equation}
The solution to the master equation (\ref{ME}) is sough for as the
expansion of the action $S_{BV}(\varphi, \varphi^*)$ w.r.t. the
resolution degree,
\begin{equation}\label{S-expansion}
    S_{BV}(\varphi,\varphi^*)=\sum_{k=0}\stackrel{(k)}{S}\, , \qquad
    deg \stackrel{(k)}{S} =k \, .
\end{equation}
Once the solution is found in all the orders resolution degree, the
complete Stueckelberg action is extracted as zero order w.r.t. to
the anti-ghost number (i.e. with switched off anti-fields), while
the Stueckelberg gauge generators are defined by the first order of
$S_{BV}$ w.r.t. the anti-ghost degree (i.e. as the coefficients at
$\xi^*_\gamma$ and $\phi^*_i$).

Consider the master action up to the next order of the resolution
degree after the boundary condition (\ref{BC-St}),
\begin{eqnarray}
 \nonumber
  S_{BV}(\varphi,\varphi*) &=& S(\phi) - \tau_\alpha \xi^\alpha + C^\alpha\left(\Gamma^i_\alpha (\phi) \phi_i^* + \xi_\alpha^*\right) +\\
 \nonumber {\phantom{S}} &+& \frac12 W_{\alpha\beta}\xi^\alpha\xi^\beta +    C^\alpha  \left(R_{\alpha\beta}^\gamma\xi^\beta \xi^*_\gamma + R_{\alpha \beta}^i\xi^\beta  \phi^*_i\right) + \\
   {\phantom{S}} &+&   \frac12 C^\beta C^\alpha
   \left( U^\gamma_{\alpha\beta} C^*_\gamma + \phi_j^*\phi^*_i E_{\alpha\beta}^{ij}+
   \xi_\mu^*\phi^*_i E_{\alpha\beta}^{\mu i} + \xi_\mu^*\xi^*_\nu E_{\alpha\beta}^{\mu\nu}\right) + \dots \,
   ,
   \label{S2}
\end{eqnarray}
where all the structure coefficients are supposed to be functions of
the original fields.  The first line in this expression is the
boundary condition (\ref{BC-St}) defined by the original action, and
by the generators $\Gamma^i_\alpha$ of the consequences
$\tau_\alpha$ included in the involutive closure of the system. The
next lines include the most general expression with the resolution
degree 2, and ghost number 0. The structure coefficient involved in
$\stackrel{(2)}{S}$ are defined by the master equation. Let us
expand the l.h.s. of the master equation w.r.t. the resolution
degree up to the first order. Notice that $\stackrel{(k)}{S}, \,
k>2$ cannot contribute to zero and first orders of the expansion, so
(\ref{S2}) is sufficient at this level
\begin{equation}\label{SS0}
    (S_{BV},S_{BV})_0=2(\Gamma^i_\alpha\partial_i
    S+\tau_\alpha)C^\alpha\equiv 0\,,
\end{equation}
\begin{eqnarray}
\nonumber
  (S_{BV},S_{BV})_1 &=& 2\xi^\gamma(\Gamma^i_\alpha\partial_i\tau_\gamma-R^i_{\alpha
    \gamma}\partial_i S-R^\beta_{\alpha
    \gamma}\tau_\beta -W_{\gamma\alpha})C^\alpha- \\
\nonumber   &- & C^\alpha C^\beta\big(\phi^\ast_i(
\Gamma^j_\alpha\partial_j \Gamma_\beta^i-\Gamma^j_\beta\partial_j
\Gamma_\alpha^i-U^\gamma_{\alpha\beta}\Gamma^i_\gamma-R^i_{\alpha
    \beta}+R^i_{\beta\alpha}-E^{ji}_{\alpha\beta}\partial_j
    S-E^{i\gamma}_{\alpha\beta}\tau_\gamma)- \\
  &-&\xi^\ast_\mu (U^\mu_{\alpha\beta} - R^\mu_{\alpha
    \beta}+R^\mu_{\beta \alpha}+E^{j\mu}_{\alpha\beta}\partial_j
    S-E^{\mu\nu}_{\alpha\beta}\tau_\nu)\big)=0\,. \label{SS1}
\end{eqnarray}
Relation (\ref{SS0}) is valid, given the gauge identity (\ref{GI}).
The first order of the master equation (\ref{SS1}) holds by virtue
of identities (\ref{W}), (\ref{GG}), (\ref{UR}) upon identification
of the structure coefficients in the expansion (\ref{S2}) with
corresponding structure functions in the mentioned identities.

As we have seen, the solution to (\ref{ME}) exists up to the second
order w.r.t. resolution degree (\ref{deg}). Let us consider the
general order $k$. Substitute the expansion (\ref{S-expansion}) into
the master equation and take $k$-th order. It has the following
structure
\begin{equation}\label{dSk}
    (S_{BV},S_{BV})_k = \delta \stackrel{(k+1)}{S}+ B_k (S, \stackrel{(1)}{S} , \dots ,
    \stackrel{(k)}{S})\, ,
\end{equation}
where $B_k$ involves only $\stackrel{(l)}{S}),\, l\leq k$, and the
operator $\delta$ reads:
\begin{equation} \label{KTd}
 \delta O=-\frac{\partial^R O}{\partial \phi^*_i}\partial_iS - \frac{\partial^R O}{\partial
    \xi^*_\alpha}\tau_\alpha +\frac{\partial^R O}{\partial
    C^*_\alpha}\left( \phi^*_i\Gamma^i_\alpha + \xi^*_\alpha\right)+\frac{\partial^R O}{\partial \xi^\alpha}C^\alpha\, .
\end{equation}
By virtue of identity (\ref{GI}), the operator $\delta$ squares to
zero,
\begin{equation} \label{KTd2}
 \delta^2 O=\frac{\partial^R O}{\partial
    C^*_\alpha}\left(\Gamma^i_\alpha\partial_i S + \tau_\alpha\right)\equiv 0
    ,
\end{equation}
 so it is a differential. Obviously, $\delta$ decreases the resolution
degree by one,
\begin{equation}\label{deg-delta}
deg (\delta)= -1 \, .
\end{equation}
Notice that $\delta$ is acyclic in the strictly positive resolution
degree because the identities (\ref{Closure}) are independent, i.e.
\begin{equation}\label{acyclic}
    \delta X =0,\, deg(X)>0  \quad\Leftrightarrow\quad \exists \,Y:\,
  X=\delta Y\,  .
\end{equation}
By Jacobi identity $(S,(S,S))\equiv 0, \, \forall S$. Expanding the
identity w.r.t. the resolution degree, one can see that $B_k$ of
relation (\ref{dSk}) is $\delta$-closed,
\begin{equation}\label{deltaB}
    \delta B_k=0, \quad k>0 .
\end{equation}
Then, because of (\ref{acyclic}), $B_k$ is $\delta$-exact,
\begin{equation}\label{B-exact}
    \exists Y_{k+1}: \,\, B_k=\delta  Y_{k+1}, \quad  deg (Y_{k+1}) = k+1
    \, .
\end{equation}
Substituting (\ref{B-exact}) into (\ref{dSk}) we arrive at the
relation
\begin{equation}\label{dS_Y}
    \delta\left(\stackrel{(k+1)}{S} + Y_{k+1}\right)=0 \, .
\end{equation}
This provides solution for $\stackrel{(k+1)}{S}$
\begin{equation}\label{Sk1}
 \stackrel{(k+1)}{S}=-  Y_{k+1} + \delta Z_{k+2},\qquad
 deg(Z_{k+2})=k+2 \, .
\end{equation}
The solution is unique modulo natural $\delta$-exact ambiguity.

In this way, one can iteratively find the master action of the
Stueckelberg theory, given the original action, generators
$\Gamma^i_\alpha$ of consequences $\tau_\alpha$ (\ref{tau}) included
into the involutive closure (\ref{Closure}) of Lagrangian system.
The solution is unobstructed at any order. Once the master action is
found, its zero order w.r.t. the anti-ghost number defines complete
Stueckelberg action, while the Stueckelberg gauge generators are
defined by the first order of $agh$, in accordance with
(\ref{BC-usual}).

In the end of this section,  notice some similarity between the BV
formalism for the Stueckelberg embedding described above, and the BV
formalism for the field theories with unfree gauge
symmetry\footnote{By unfree gauge symmetry, we mean the case when
the gauge parameters are constrained by differential equations. The
most known example of the unfree gauge symmetry is the unimodular
gravity where the gauge parameters are constrained by transversality
equation.} \cite{Unfree-BV}. In the latter case, the BV formalism
\cite{Unfree-BV} also involves the compensatory fields $\xi$ with
ghost number $0$ and resolution degree $1$.  These fields
compensate, in a sense, the constraints imposed on the gauge
parameters making the effective gauge symmetry parameters
unconstrained. These variables can be viewed, in a broad sense, as
Stueckelberg fields. In the case of unfree gauge symmetry, however,
there is no pairing between the gauge symmetry parameters and the
compensator fields, unlike the case of Stueckelberg symmetry
considered in this article.

\section{Concluding remarks}
Let us make concluding remarks and discuss further perspectives. At
first, let us sum up the proposed scheme of including Stueckelberg
fields.

Proposed inclusion of Stueckelberg fields proceeds from involutive closure of Lagrangian system (\ref{Closure}) which includes the consequences (\ref{tau}) of the original Lagrangian equations. The involutive system (\ref{Closure})  does not admit any lower order consequence, while the original Lagrangian equations are not necessarily involutive. The Hamiltonian analogue of the involutive closure is the completion of the original equations by secondary constraints, being zero order consequences of the original primary constraints and the first order Hamiltonian equations.
The involutive closure of Lagrangian equations allows explicitly covariant degree of freedom count,
see relations (\ref{DoF0}), (\ref{DoF1}) in the Appendix.
The explicit control of the DoF number in the involutive closure allows one to consistently include interactions in the covariant form for the second class systems, see (\ref{Vertices}), (\ref{Gamma-vertices}), (\ref{tau-vertices}). This method has been first proposed in the article \cite{Involution} (for applications of the method in specific models see, for example, \cite{Cortese:2013lda}, \cite{Kulaxizi:2014yxa}, \cite{Abakumova:2018eck}, \cite{Rahman:2020qal}). The original proposal of \cite{Involution} allows one to find all the consistent vertices in the field equations, including non-Lagrangian one, without distinctions between variational and non-variational interactions. In this articles we added a side remark which allows one to identify all the consistent Lagrangian interactions. In this article, however, the involutive closure is considered not for its own sake, but as a starting point for inclusion of Stueckelberg fields.

The proposed procedure implies to introduce the  Stueckelberg field
$\xi^\alpha$ for every consequence (\ref{tau}) added to the
Lagrangian equations to form the involutive closure (\ref{Closure}).
The Stueckelberg action (\ref{St-action}) and gauge symmetry
generators (\ref{gt}) are  sought for as the power series in
Stueckelberg fields proceeding from the requirement of gauge
symmetry.  The boundary conditions (\ref{first-approx}) for the
action and gauge generators are defined by the original action and
generators $\Gamma^i_\alpha$ of consequences $\tau_\alpha$ included
into the involutive closure (\ref{Closure}). Given the boundary
conditions, relations (\ref{NI-k}), being the expansion coefficients
of the Noether identities for Stueckelberg gauge symmetry, allow one
to iteratively find order by order all the expansion coefficients of
the action and gauge generators. This iterative procedure of
inclusion Stueckelberg fields is unobstructed at every stage. Upon
inclusion of Stueckelberg fields, the theory has the same DoF number
(see in the Appendix), while the gauge fixing $\xi=0$ is admissible.
This means the equivalence of the Stueckelberg theory to the
original one. It is worse to mention that existence of the
Stueckelberg embedding of the original system does not impose any
restriction on the generators $\Gamma^i_\alpha$ (\ref{tau}) besides
independence of consequences $\tau_\alpha$  included into involutive
system (\ref{Closure}. In particular, the generators of consequences
$\Gamma$, being the leading terms in Stueckelberg gauge symmetry
transformations of original fields (\ref{first-approx}), are not
required to form the integrable distribution, even on shell. This
contrasts to the logic of the most common form of "Stueckelberg
trick" which implies that the Stueckelberg gauge transformations of
the original fields begin with the gauge generators of gauge
symmetry of a part of original action. Being the gauge generators,
they inevitably commute to each other, at least on shell. As is seen
from this article, the integrability of the distribution generated
by $\Gamma$'s is unnecessary for existence of consistent
Stueckelberg embedding. The algebraic structure behind the
consistency of the proposed Stueckleberg embedding is described in
Section 3. This algebra follows from the gauge identities (\ref{GI})
of the involutive closure (\ref{Closure}) much like the gauge
algebra of the gauge invariant system follows from the Noether
identities. In the case of Noether identities of gauge invariant
system, the structure relations of the gauge algebra are deduced
proceeding from the three factors: (i) dependence of the equations,
(ii) independence of the generators of identities, and (iii) the
gauge symmetry of the equations. It is the third factor which does
not apply to the gauge identity (\ref{GI}), while the first two ones
do. Therefore, similar structure relations follow from the first two
factors, while the integrability of the gauge distribution, being a
consequence of the third factor is not required. The structure
relations of the gauge algebra of the closure of Lagrangian system,
being described in Section 3, are equivalent to the equations of
Section 2 defining the Stueckelberg action and Stueckelberg gauge
gauge generators. Once the gauge algebra is consistent (as it is a
consequence of the non-contradictory identities (\ref{GI})),  this
explains why the Stueckelberg embedding is unobstructed at all the
stages.

In this article, also BV formalism is proposed to perform the
Stueckelberg  embedding proceeding from the involutive closure of
Lagrangian system (\ref{Closure}). There are three key distinctions
from the conventional BV-formalism \cite{BV1}, \cite{BV4}. The first
difference is that the boundary conditions (\ref{BC-St}) imply to
specify, besides the action, also the generators of consequences
$\Gamma^i_\alpha$ and the consequences $\tau_\alpha$ included into
the involutive closure of Lagrangian equations (\ref{Closure}). The
second difference is the set of variables. Once the original action
is non-gauge invariant, no ghost would be introduced under the usual
BV procedure \cite{BV1}. For inclusion of the Stueckelberg
embedding, the ghosts are assigned to every generator of consequence
$\Gamma^i_\alpha$, and the Stueckelberg field $\xi^\alpha$ is
introduced for every consequence $\tau_\alpha$ (\ref{grad-fields}).
The anti-fields are introduced for all the fields, including ghosts
and Stueckelberg fields. The third distinction is that the solution
of the master equation is iterated w.r.t. a different resolution
degree (\ref{deg}) which counts the orders both anti-fields and
Stueckelberg fields. The solution of master equation exists at every
order of the iterative procedure, no obstruction can arise. Once the
master action is found, the anti-field independent part gives the
Stueckelberg action, while the the first order in the anti-fields
gives the generators of Stueckelberg gauges symmetry. This gives
another way to construct the Stueckelberg embedding of the original
theory.

Concerning further perspectives, two directions can be mentioned for
developing the proposed method. The first direction is related to
the possibilities of applying this method to specific models of
field theory. From this perspective the potential advantage of the
method is that it allows one to control the degree of freedom number
in an explicitly covariant fashion at every stage, be it inclusion
of interactions, on Stueckelberg embedding. Among the models of
current interest various generalizations of gravity can be mentioned
where the original Lagrangian equations are not involutive, and
hence the second class constraints arise. On the other hand, in
these models do not offer any obvious hint for the ``Stueckelberg
trick'' which would correspond to the conversion of the second class
constraints at Hamiltonian level. In particular, this concerns
generalizations of unimodular gravity \cite{Barvinsky:2017pmm},
\cite{Barvinsky:2019agh} and ``new massive gravity" in $3d$
\cite{Bergshoeff:2009hq}, \cite{Bergshoeff:2009aq},
\cite{Ozkan:2018cxj}. The proposed method seems an appropriate tool
to study dynamics in the models of this type. These models, being
complex by themselves, are not suitable, however, as touchstones for
the first article no the general method, so the applications will be
studied elsewhere.

Another aspect of possible future developments concerns
generalizations of the method as such. With this regard, the
inclusion of the case of gauge invariant actions whose Lagrangian
equations admit the lower order consequences. The extension of the
procedure for inclusion Stueckelberg fields seems straightforward.
One more issue concerns the case where the reducible set of
consequences is included into the involutive closure (\ref{Closure})
of Lagrangian system. Notice that the involutive closure is not
unique, as the higher order consequences can be added without
breaking involutivity, and also the generating set of consequences
can be chosen over-complete. Therefore, the same action can admit
different involutively closed extensions of Lagrangian equations.
The set of Stueckelberg fields is defined by specific involutive
closure of Lagrangian equations. That is why, the multiple choice is
possible for the set of Stueckelberg fields for the same Lagrangian,
including, in principle, reducible sets of gauge generators. One
more potential aspect of interest in this story is related to gauge
fixing. Hypothetically, there can be the case when the gauge fixing
is admissible which kills the original fields while the physical
degrees of freedom are described by the Stueckelberg fields. This
would allow to construct dual formulations of the same dynamics
connecting them by different schemes of inclusion Stueckelberg
fields and different gauge fixing.

Finally, let us mention the issue of locality of the Stueckelberg
embedding procedure proposed in this article.  The existence of the
embedding is proven in Section 4 in terms of condensed notation that
could potentially hide the obstructions related to locality. There
is, however, a non-rigorous reason to believe that the locality
problem should not arise. Once the action and the generators of
consequences (\ref{tau}) are of the finite order, the structure
functions of the algebra (see Section 3), being consequences of the
identities (\ref{GI}), should involve the finite number of the
derivatives. The theory involve a natural invertible operator
$W_{\alpha\beta}$ (\ref{W}) which defines the kinetic term for
Stueckelberg fields. The inverse can be non-local, but the embedding
procedure does not need to invert $W$.

\subsection*{Acknowledgements} The author thanks  A.~Sharapov for valuable discussions of this work and comments on the manuscript.
The part of the work concerning inclusion of Stueckelberg fields is supported by the Foundation for the Advancement of Theoretical
Physics and Mathematics ``BASIS". The BV-formalism for inclusion of Stueckelberg fields is worked out with the support of the
Ministry of Science and Higher Education of the Russian Federation, Project No. 0721-2020-0033.

\pagebreak

\section*{ Appendix  \\
 Degree of freedom count in involutive closure of
Lagrangian systems} In this Appendix, the relations are provided for
DoF number count in the involutively closed systems. First these
relations have been obtained in the article \cite{Involution} for
general involutive field equations\footnote{Certain regularity
assumptions are made for deducing these relations, see in
\cite{Involution}. Here, the regularity issues are not addressed.},
not necessarily being the involutive closure of any Lagrangian
system. Here, these relations are slightly re-arranged to be more
convenient for making simplifications related to the systems being
the involutive closure of Lagrangian equations (\ref{Closure}).

In the Appendix, all the indices are  by default understood as numerical labels, not
condensed ones. Exceptions are specially reported in each case.

The degree of freedom number is counted in terms of orders of equations, gauge identities and gauge symmetries.
Let us explain the definitions of these orders.

Consider a system of field equations
\begin{equation}\label{IE}
    E_I(\phi, \partial\phi,\dots, \partial^{n_I}\phi)=0\, ,
\end{equation}
where $n_I$ is the maximal order of partial derivatives involved
in the equation with the label $I$. The number $n_I$ is considered as the order of equation $n_I$.
The partial derivatives by all
space-time coordinates are treated on the equal footing, and the
order $n_I$ accounts for the mixed derivatives. For example, the order
of the equation $\frac{ \partial^2 \phi}{\partial x \partial t}=0$ is
2.

A system of equations is considered involutive if it does not admit such consequences
of a lower order that are not yet included in the original system.
If the system is not involutive, it can be
always complemented by the  lower order consequences, to make it
involutive. In principle, higher order consequences can be also
added, to make the system involutive. Completion of the system by the consequences is understood as
involutive closure of the system, if no new lower order consequences exist.
The involutive closure has the same solutions as the original system.
In this sense, the involutive closure is equivalent to the original system.

Let the  equations (\ref{IE}) admit gauge identities
\begin{equation}\label{IGI}
L_A^I E_I\equiv 0\, ,
\end{equation}
where $L_A^I$ are the differential operators with the field
depending coefficients,
\begin{equation}\label{L-op}
L_A^I=\sum_{k=0}^{l_A^I}
L_A^I{}^{\mu_1\dots\mu_k}(\phi,\partial\phi,
\partial^2\phi\dots)\partial_{\mu_1}\dots\partial_{\mu_k}\, \qquad
l_A^I\in\mathbb{N}.
\end{equation}
Operator $L_A^I$ is called the generator of gauge identity, and
$l_A^I$ is the order of the operator. The order $i_A$ of the gauge
identity (\ref{IGI}) with the label $A$ is defined by the following
rule:
\begin{equation}\label{oId}
i_A=\max\limits_{\{I\}} \left(l^I_A+n_I\right) \, ,
\end{equation}
i.e. it is the maximal aggregate order of the identity generator and
the equation it acts on.

Suppose the involutive system (\ref{IE}) admits gauge symmetry
transformations,
\begin{equation}\label{rgst}
\delta_\epsilon\phi^i= R^i_\alpha \epsilon^\alpha \,, \qquad
\delta_\epsilon E_I\approx 0 \,, \quad \forall\epsilon^\alpha \, ,
\end{equation}
where the gauge generators $R^i_\alpha$ are the
differential operators with the field-depending coefficients,
\begin{equation}\label{Rgenerator}
R^i_\alpha =\sum_{k=0}^{r^i_\alpha}R^{i\mu_1
\dots\mu_k}_\alpha(\phi,\partial\phi, \dots
)\,\partial_{\mu_1}\cdots\,\partial_{\mu_k} .
\end{equation}
The gauge variation (\ref{rgst}), by definition, leaves the mass shell invariant.

The order $r_\alpha$ of the gauge transformation with specific
parameter $\epsilon^\alpha$ is defined as the maximal number of
derivatives acting on the parameter in the transformation of any
field,
\begin{equation}\label{gtorder}
   r_\alpha=\max\limits_{\{i\}} \left(r^i_\alpha\right) \, .
\end{equation}
Given the involutive equations with the complete set of gauge
identities, and gauge symmetries, the DoF number is counted by the rule
\begin{equation}\label{DoF0}
    N_{DoF}=\sum_{I}n_I-\sum_{A}i_A - \sum_{\alpha}r_\alpha
\end{equation}
So, the DoF is computed as follows: the total order of the
identities and the total order of gauge symmetries are subtracted
from the total order of the equations. Notice two interesting
features of this counting recipe. First, it does not involve the
number of fields. Second, zero orders relations (of any sort, be it
equation, identity, or gauge symmetry) do not contribute to the DoF
number. This relation for $N_{DoF}$ has been first deduced in the
article \cite{Involution} in a slightly different form. The recipe
(\ref{DoF0}), being equivalent to the original one, is more
convenient for simplifications accounting for specifics of
involutive closure of Lagrangian systems.

Let us apply the DoF number counting recipe (\ref{DoF0}) first to the involutive closure of Lagrangian system (\ref{Closure}).
Denote as  $n_\alpha$ the order of consequence $\tau_\alpha$ (\ref{tau}), and let $g_\alpha^i$ be the order of the differential operator $\Gamma^i_\alpha$ generating the consequence, and $n_i$ is the order of Lagrangian equation (\ref{EoM}),
By construction (\ref{tau}),  the order of consequence $\tau_\alpha$ cannot exceed the maximal
 aggregate order of the generator of consequence and  original Lagrangian equations (\ref{EoM})
 \begin{equation}\label{nalpha}
n_\alpha\leq \max\limits_{\{i\}}\left(g_\alpha^i+n_i\right)\, .
 \end{equation}
If only lower-order consequences are included, then this is a strict inequality.
With the consequences whose order is higher than the original equations, the equality is possible.
The maximum of the l.h.s. inequality (\ref{nalpha}) is reached at certain $i$, which is denoted $\bar{i}$,
\begin{equation}\label{bari}
\max\limits_{\{i\}}\left(g_\alpha^i+n_i\right)=g_\alpha^{\bar{i}}+n_{\bar{i}} \, .
\end{equation}
The order of the corresponding operator $\Gamma^{\bar{i}}_\alpha$ is unique for given $\alpha$.
This order is denoted just $g_\alpha$.
In the identity (\ref{GI}), the coefficient at $\tau_\alpha$ has zero order. Given the inequality (\ref{nalpha}), the order of identity (\ref{GI}), being defined by the rule (\ref{oId}), reads
\begin{equation}\label{InIdO}
 i_\alpha= g_\alpha+n_{\bar{i}} \, .
\end{equation}
As a result, for the DoF number of any Lagrangian system without gauge symmetry\footnote{If the Lagrangian system had gauge symmetry, one should additionally subtract the total order of gauge symmetry transformations, see (\ref{Rgenerator}), (\ref{gtorder})} can be counted making use of the orders related to its involutive closure (\ref{Closure}):
\begin{equation}\label{DoF1}
    N_{DoF}=\sum_i n_i+ \sum_\alpha (n_\alpha -i_\alpha)= \sum_{i\neq\bar{i}} n_i + \sum_\alpha (n_\alpha -g_\alpha)\, .
\end{equation}
The above formula allows one to count DoF number of any Lagrangian
system in explicitly covariant way. In this form, it works for the
case without gauge symmetry (second class constrained systems, from
Hamiltonian perspective). To account for the gauge symmetry, one has
to subtract the total order of gauge symmetry generators (see
(\ref{DoF0})) in the irreducible case. Further adjustments can be
made for the reducible gauge symmetry.

Now, let us discuss the DoF number upon inclusion Stueckelberg fields.
Consider the Stueckelberg action and gauge symmetry transformations,\footnote{Here, the labels are understood as condensed indices.}
\begin{equation}\label{Stueck-Lead}
  \mathcal{S}_{St}=S(\phi)+\tau_\alpha\xi^\alpha+ \frac12
  W_{\alpha\beta}\xi^\alpha\xi^\beta+\cdots\, , \qquad \delta_\epsilon\xi^\alpha=\epsilon^\alpha+\cdots\,,\quad\delta_\epsilon\phi^i=\Gamma^i_\alpha\epsilon^\alpha+\cdots ,
  \end{equation}
where $\cdots $ stand for the higher orders in $\xi$. From the perspective of the DoF number counting, there are two main distinctions between the involutive closure  (\ref{Closure}) of the original Lagrangian system, and the Stueckelberg theory (\ref{Stueck-Lead}).
First, the consequences $\tau_\alpha$  are replaced by Lagrangian equations for $\xi^\alpha$
\begin{equation}\label{xi-eq}
  \tau_\alpha(\phi) \quad\mapsto\quad\frac{\partial \mathcal{S}_{St}}{\partial\xi^\alpha}=\tau_\alpha+W_{\alpha\beta}\xi^\beta + \cdots\, ,
\end{equation}
where the indices are condensed.  Denote $\bar{n}_\alpha$ the order of the equation for Stueckelberg field $\xi^\alpha$
Notice that the operator $W_{\alpha\beta}$ results from the action of the operator
$\Gamma^i_\alpha$ on $\tau_\beta$. This means that the total order of the equations $\frac{\partial \mathcal{S}_{St}}{\partial\xi^\alpha}$ exceeds the total order of the equations $\tau_\beta$ by the total order of operators $\Gamma$, i.e.
\begin{equation}\label{nbar}
  \sum_\alpha\bar{n}_\alpha=\sum_\alpha\left(n_\alpha+g_\alpha\right).
\end{equation}
Hence, inclusion of Stueckelberg fields increases the positive contribution to the DoF number (\ref{DoF1}) by $\sum_\alpha g_\alpha$.
The second relevant distinction of the Stueckelberg theory (\ref{Stueck-Lead}) from the involutive closure of original Lagrangian equations (\ref{Closure}) is the gauge symmetry. The order of the gauge transformation is $g_\alpha$, at least in the first approximation in Stueckelberg fields. Hence, the negative part of the DoF count changes to the same number $\sum_\alpha g_\alpha$. These two changes cancel each other, so the DoF number remains unchanged.

Below, we illustrate involutive closure, inclusion of Stueckelberg fields and DoF counting by two simple examples: mechanical toy model, and Proca Lagrangian.

\subsection*{Example 1. Mechanical toy model}
As a toy example, consider $L=\frac12\dot{x}{}^2 $.
The Lagrangian equation, $\ddot{x}=0$,
is involutive, as it does not admit any
lower order differential consequence.
However, one could add a higher differential consequence, being just a derivative of the equation.
This extension is also involutive system, as no lower order consequences exist.
The system (\ref{Closure}) in this case reads,
\begin{equation}\label{Clos1}
  \frac{\delta S}{\delta x} =-\ddot{x} =0\, ,\qquad \tau=\dddot{x}=0\,,
\end{equation}
and the generator of the consequence $\Gamma$ is just $\frac{d}{dt}$.
The identity (\ref{GI}) between the equations (\ref{Closure}) in this case reads
\begin{equation}\label{GI1}
  \tau +\Gamma\frac{\delta S}{\delta x}\equiv \dddot{x} - \dddot{x}\equiv 0
\end{equation}
Let us exemplify the DoF count formula (\ref{DoF1}) by the toy model (\ref{Clos1}). There is one Lagrangian equation of the second order $\ddot{x}$, i.e. $n_i=2$. There is one consequence of the third order $\tau=\dddot{x}$, i.e. $n_\alpha=3$. There is one generator of consequence $\Gamma=\frac{d}{dt}$,  it is of the first order, i.e. $g=1$. Substituting these numbers into (\ref{DoF1}), one gets $N_{DoF}=2+3 -(2+1)=2$, as it should be.
Now, given the involutive system of Lagrangian equation and its consequence (\ref{Clos1}), let us include the Stueckelberg fields following the procedure of Section 2. The operator $W$ (\ref{W}) in this case is just $\frac{d^4}{dt^4}$, so the Stueckelberg Lagrangian (modulo total derivative) and gauge symmetry transformations (\ref{Stueck-Lead}) read
\begin{equation}\label{L-e1}
  L_{St}=\frac12\left(\dot{x}+\ddot{\xi}\right)^2\,,\qquad \delta_\epsilon\xi=\epsilon\, , \quad \delta_\epsilon x= -\dot{\epsilon} \, .
\end{equation}
The degree of freedom count adds one to the positive part of the sum (\ref{DoF1}) as the third order consequence $\dddot{x}=0$ is replaced by the fourth order equation for $\xi$. Simultaneously, 1 is added to the negative contribution, because the first order gauge symmetry is switched on.
Obviously the Noether identity for Lagrangian (\ref{L-e1}) at $\xi=0$ reproduces the identity (\ref{GI1}) of the system (\ref{Clos1}).
The equivalence  of the Stueckelberg Lagrangian to the original one is obvious, as the gauge fixing condition $\xi=0$ is admissible, and the Stueckelberg equations reduce to (\ref{Clos1}) in this gauge.

Let us comment on this elementary example from the perspective of conversion method for the Hamiltonian second class constrained systems. The Lagrangian $L=\frac12\dot{x}{}^2$ could be considered as including accelerations. Then, the Ostrogradski method should be applied. The velocity would become an auxiliary canonical coordinate, whose conjugate momentum should vanish due to the primary constraint. Conservation of the primary constraint results in the secondary constraint which connects the canonical momentum of the original coordinate with the velocity. The pair of primary and secondary constraints is of the second class. If they are converted into the first class, we arrive at gauge invariant Hamiltonian action. All the momenta can be excluded by the inverse Legendre transformation in this action, and we arrive at Lagrangian (\ref{L-e1}). This example demonstrates that the covariant procedure of inclusion Stueckelberg fields proposed in Section 2 directly corresponds to the Hamiltonian conversion of the second class systems. It also illustrates the fact that the method works well proceeding from any extension of Lagrangian system by the consequences, including the higher order ones, if the starting point is an involutively closed system.

\subsection*{Example 2. Proca action}
Let us exemplify DoF count recipe (\ref{DoF1}) by the Proca
equations for massive vector field,
\begin{equation}\label{Proca-eq}
\frac{\delta S_{Proca}}{\delta A_\mu}\equiv\left(\delta^\mu_\nu(\Box
- m^2) -
\partial^\mu\partial_\nu\right)A^\nu=0\, .
\end{equation}
The Proca equations is a system are of four equations of the second
order, so $n_i=2, i=1,2,3,4$. Therefore, the total order of Lagrangian equations is $8$.   The Proca equations are not involutive as they
admit the first order consequence:
\begin{equation}\label{tau-Proca}
\tau=-\partial^\mu\frac{\delta S_{Proca}}{\delta A^\mu}\equiv m^2
\partial_\mu A^\mu\, .
\end{equation}
So, we have one first order consequence $\tau$, to be added for the
sake of involutive closure. This means, $n_\alpha=1$. The generator $\Gamma^\mu=\partial^\mu$ of the single consequence is the operator of divergence. It has the order one, $g=1$. No other equations
and consequences are included in the involutive closure
(\ref{Proca-eq}), (\ref{tau-Proca}) of Proca system.  The gauge identity (\ref{GI}) for the Proca system reads,
\begin{equation}\label{GI-Proca}
    \partial^\mu \frac{\delta S_{Proca}}{\delta A^\mu}+\tau\equiv 0 \,
    .
\end{equation}
The identity is of the order $3$, as the first order operator $\partial$ acts on the second order Proca equations.
Now, let us calculate the DoF number applying relation (\ref{DoF1}).
The total order of the equations in the involutive closure of the Proca system is $9$:
there are $4$ second order Proca equations  (\ref{Proca-eq}), plus one first order consequence,
$\sum_i n_i+\sum_\alpha n_\alpha= 4\times 2+1=9$.
There is also one third order gauge identity (\ref{GI-Proca}),
$\sum_\alpha i_\alpha=3$. Substituting these numbers into the formula (\ref{DoF1}), we get $9-3=6$.
So it is 6 DoF's by the phase space count, that corresponds to 3 by configuration space count, as it should be for the massive spin 1 in $d=4$.

Given the involutive closure of the Proca equations, let us
introduce the Stueckelberg field following the prescription of
Section 2. The matrix $W$ (\ref{W}) in this case is the
d'Alembertian, and no higher order corrections appear. So we arrive
to the standard Stueckelberg equations, gauge symmetries, and
Noether identities
\begin{equation}\label{Stueck-eq}
    \frac{\delta \mathcal{S}}{\delta\xi}= m^2(\partial_\mu A^\mu
    -\Box\xi)=0, \qquad \frac{\delta \mathcal{S}}{\delta A_\mu}=\left( \delta^\mu_\nu \Box
    -\partial_\nu
    \partial^\mu\right)A^\nu + m^2(A_\mu-\partial_\mu\xi)=0;
\end{equation}
\begin{equation}\label{gstProca}
    \delta_\epsilon\xi=\epsilon\,, \quad \delta_\epsilon
    A_\mu=\partial_\mu\epsilon\, ; \qquad \frac{\delta
    \mathcal{S}}{\delta\xi}-\partial_\mu \frac{\delta \mathcal{S}}{\delta
    A_\mu}\equiv 0\, .
\end{equation}
Let us apply to this system the DoF number count recipe
(\ref{DoF0}). There are five second order equations
(\ref{Stueck-eq}), so the total order of the equations is ten. There
is one first order gauge symmetry, and one third order gauge
identity. So, the DoF number is $10-3-1=6$, as it should be.

\pagebreak

\end{document}